\documentclass[a4paper,11pt]{article}
\usepackage{jheppub}

\newcommand{\oarX}[1]{\href{http://arxiv.org/abs/#1}{{\ttfamily #1}}}
\newcommand{\arX}[1]{\href{http://arxiv.org/abs/#1}{{\ttfamily arXiv:#1}}}

\def\barr{\begin{array}}
\def\earr{\end{array}}

\def\ben{\begin{equation}}
\def\een{\end{equation}}
\def\bs{\begin{subequations}}
\def\es{\end{subequations}}
\def\bena{\begin{eqnarray}}
\def\eena{\end{eqnarray}}

\def\bR{\mathbb{R}}
\def\bC{\mathbb{C}}

\def\O{{\rm O}}

\def\SU{{\rm SU}}
\def\SL{{\rm SL}}

\def\im{{\rm i}}

\newcommand{\ie}{\textit{i.e.}}
\newcommand{\eg}{\textit{e.g.}}

\title{\boldmath Identifying cosmological perturbations in group field theory condensates}

\author{Steffen Gielen}
\affiliation{Theoretical Physics, Blackett Laboratory, Imperial College London \\ London SW7 2AZ, United Kingdom}

\emailAdd{sgielen@imperial.ac.uk}

\abstract{One proposal for deriving effective cosmological models from theories of quantum gravity is to view the former as a mean-field (hydrodynamic) description of the latter, which describes a universe formed by a `condensate' of quanta of geometry. This idea has been successfully applied within the setting of group field theory (GFT), a quantum field theory of `atoms of space' which can form such a condensate. We further clarify the interpretation of this mean-field approximation, and show how it can be used to obtain a semiclassical description of the GFT, in which the mean field encodes a classical statistical distribution of geometric data. In this sense, GFT condensates are quantum homogeneous geometries that also contain statistical information about cosmological inhomogeneities. We show in the isotropic case how this information can be extracted from geometric GFT observables and mapped to quantities of observational interest. Basic uncertainty relations of (non-commutative) Fourier transforms imply that this statistical description can only be compatible with the observed near-homogeneity of the Universe if the typical length scale associated to the distribution is much larger than the fundamental `Planck' scale. As an example of effective cosmological equations derived from the GFT dynamics, we then use a simple approximation in one class of GFT models to derive the `improved dynamics' prescription of holonomy corrections in loop quantum cosmology.}

\keywords{Models of Quantum Gravity, Cosmology of Theories beyond the SM, Lattice Models of Gravity}

\begin{document} 
\maketitle
\flushbottom

\section{Introduction}
\label{sec:intro}

The most promising hope for connecting theories of quantum gravity with observation is to understand their consequences for early universe cosmology. Conversely, input from quantum gravity may be needed in order to complete or replace the standard paradigm of inflation. Open theoretical issues of inflation, such as the choice of initial conditions or the origin of the inflationary potential \cite{steinh} or the need for a resolution of the past singularity of inflationary universes \cite{borde}, should ultimately be addressed in a more complete framework. More generally, with a deeper understanding of quantum gravity, one might obtain a different physical picture for the very early universe that does not require introducing scalar fields, but is at the same time compatible with observations such as made by Planck \cite{planck}. 

In order to obtain clear predictions from quantum gravity-inspired approaches to cosmology, it is essential to understand their precise relation to an underlying quantum gravity theory. One such approach is loop quantum cosmology (LQC) \cite{LQC1,LQC2,LQC3}, in which the classical Big Bang singularity is resolved \cite{bounce} and the issue of initial conditions for inflation can be addressed \cite{LQCinfl}, but whose relation to the full theory of loop quantum gravity (LQG) \cite{LQG1,LQG2,LQG3}, or the closely related group field theory (GFT) \cite{GFT1,GFT2}, a {\em second quantised} version of LQG \cite{daniele2q}, is not fully clear. In LQC, a symmetry reduction to a minisuperspace model is performed before quantising the remaining degrees of freedom with LQG techniques, leading to quantisation ambiguities and obscuring the potential embedding into the full theory.

One perspective on addressing this issue, advocated in \cite{martinreview}, is to view LQC and minisuperspace quantum cosmology as `single-patch theories' in which an elementary small chunk of space is quantised in order to capture the dynamics of an exactly homogeneous universe. One expects a more complete picture, rich enough to also capture inhomogeneities, to arise from a `many-patch theory' in which many such chunks can interact. Depending on the physical length scales one associates with these chunks, one might think of this as related to the separate-universe approach in cosmology \cite{sepuniv}. This perspective then calls for an application of concepts and methods from many-body quantum systems in condensed matter physics to quantum cosmology, as it suggests thinking of a macroscopic universe as a {\em condensate} of many such `atoms' of space. This idea had been discussed in various contexts, from the perspective of analogue gravity \cite{hu} as well as in quantum gravity \cite{danielecond1,danielecond2}. It was then explored in \cite{nonlin} in a lattice model starting from classical general relativity.

The viewpoint that a cosmological universe arises from the condensation of many `atoms of space' is most naturally investigated in the setting of group field theory (GFT), which provides a quantum field theory language for discrete (simplicial) geometry in which the concept of a {\em condensate} can be made sense of: a GFT condensate defines a non-perturbative ground state, describing a phase away from the Fock vacuum around which perturbative physics is defined in LQG. This Fock vacuum is analogous to the Ashtekar-Lewandowski vacuum \cite{ashlew} of LQG, and the excitations around it form the vertices of LQG spin networks. This vacuum corresponds to a degenerate geometry (zero expectation value for areas, volumes etc), and a non-degenerate continuum must be sought away from it.

The idea of deriving quantum cosmology models from the dynamics of a GFT condensate was explored in a series of papers \cite{prl,condlong,example,nscaling} (see also the related work \cite{gianl,lor,naivepert,newcon}). In \cite{prl,condlong}, approximating a physical state with a (generalised) coherent state of the GFT field operator, and hence working, in the simplest case, in a mean-field approximation, it was shown that Schwinger-Dyson equations encoding the GFT dynamics reduce to nonlinear, nonlocal differential equations for the mean field, or `condensate wavefunction'. Furthermore, in the simplest approximation, one obtains a {\em linear} equation that resembles a Wheeler-DeWitt equation in standard quantum cosmology. In a WKB limit for the isotropic case, one can reproduce the Friedmann equation of classical vacuum GR. One might view these results as suggesting a {\em direct} derivation of quantum cosmology models from the dynamics of suitable GFT Fock states, reproducing exactly the usual formalism in terms of a `wavefunction of the universe' annihilated by a Hamiltonian constraint operator \cite{qcrev}.

However, it was clear that this most direct equivalence of the effective condensate dynamics and a minisuperspace quantum theory cannot be expected. The `condensate wavefunction' is not an actual wavefunction; one can think of it as a classical field, whose amplitude and phase correspond, in the case of a Bose-Einstein condensate, to classical properties such as the density and velocity of the condensate \cite{becbook}. Its dynamics represent the collective, hydrodynamic description of the condensate. The equations defining these dynamics are nonlinear, both for real condensates (\eg,\, the Gross-Pitaevskii equation) and for GFT condensates, and so one would have to make sense of nonlinear quantum mechanics. It was already understood in \cite{example,nscaling} that the WKB approximation for the resulting dynamics does not capture a physically meaningful limit, even though it formally corresponds to $\hbar\rightarrow 0$: in a Bose-Einstein condensate, in this limit one would consider a fluid with almost constant density but very high velocity; for GFT condensates, one would assume that the individual `atoms' are very large and behave semiclassically individually. 

In this paper, we further explore and clarify the interpretation of GFT condensates in cosmological terms as outlined in \cite{nscaling}, using not a standard (`first-quantised') Wheeler-DeWitt equation or a quantum cosmology wavefunction but focussing instead on {\em expectation values} of Fock space operators that are meaningfully defined on any GFT many-particle state. These expectation values are given a cosmological interpretation which is used to obtain cosmological dynamics from dynamical relations between these expectation values. We arrive at an interpretation in which, while GFT condensates are homogeneous as quantum states, their effective classical (statistical) description in the hydrodynamic approximation
generally includes inhomogeneities. In this hydrodynamic picture, the single-particle wavefunction of the `ground state' that all quanta are condensed into is reinterpreted as a statistical probability distribution. For this distribution to define an exactly homogeneous classical geometry, one would have to demand that condensation occurs into an extremely peaked state, analogous to a delta distribution in position or momentum space for a Bose-Einstein condensate.\footnote{This sense of exact homogeneity can at best be satisfied for either metric or connection variables, due to uncertainty relations for canonically conjugate variables. For the purposes of this paper, we will focus on the notion of homogeneity in the reconstructed classical {\em metric}.} This requirement is an {\em additional} condition on top of condensation, which cannot be satisfied in general, as it depends on the details of the dynamics.

In general, the mean field has a finite spread on minisuperspace, and thus the hydrodynamic approximation leads to a statistical distribution of classical `patches' with different geometric data. Global information about the inhomogeneities, corresponding to moments $\int {\rm d}^3x\;\psi(\vec{x})^n$ in the approximate continuum description, can then be extracted from `global' expectation values of GFT operators. These moments are independent for different $n$, and knowing them for many different $n$ allows (in principle) the reconstruction of the inhomogeneities $\psi(\vec{x})$ (in the simplest case of only isotropic scalar perturbations)\footnote{Concretely, given $n$ such integrals, one chooses a suitable $n$-dimensional function space with basis $\{f_n\}$ and reconstructs the $\alpha_n$ in $\psi(\vec{x})=\sum_n\alpha_nf_n(\vec{x})$; a possible choice is, \eg, $f_n(\vec{x})=\sin(nk_0|\vec{x}|)/(nk_0|\vec{x}|)$.}. We give an example in which, in the homogeneous case, the improved dynamics prescription for holonomy corrections in LQC \cite{improv} can be obtained from such expectation values. In general, the resulting equations depend on the inhomogeneities. We also show that, as there is a fundamental limit on the minimal spread of the mean field in metric variables, a statistically almost homogeneous classical metric (\ie\, a very sharply peaked statistical distribution on minisuperspace) can only arise if the average areas in this distribution are {\em large} compared to the fundamental `Planck' scale set by the parameter $\kappa$ of the GFT. Explaining why this is the case is then the restatement in GFT condensate cosmology of the puzzle of how to explain the smallness of inhomogeneities in our own Universe. This interpretation also explains the shortcomings of the analysis of \cite{naivepert} in which it was suggested to add inhomogeneities by hand and interpret the resulting `condensate wavefunction' directly as a wavefunction for background and inhomogeneities {\em \`a la} Halliwell-Hawking \cite{hallhawk}. The hydrodynamic approximation already contains statistical inhomogeneities.

\section{Cosmology with group field theory (GFT) condensates}
\label{sec:condcosm}

This section provides a self-contained overview of GFT, condensate states in GFT and their relation to quantum cosmology. For more details on the general GFT formalism see \eg\, \cite{GFT1,GFT2}; full details of the construction of GFT condensates are given in \cite{condlong}.

Group field theory (GFT) provides a quantum field theory language for simplicial geometry and for the kinematics and dynamics of LQG. It can also be viewed as a generalisation and extension of matrix models which achieve a definition of $2d$ quantum gravity in terms of random matrices \cite{matrmod}. In the formulation used so far in the construction of GFT condensates, one uses a complex scalar field
\ben
\varphi:G^4/G\rightarrow\mathbb{C}
\een
where $G$ is a Lie group fixed from the outset, which will become the gauge group of gravity. For models of $4d$ quantum gravity, the conventional choices are $G=\SL(2,\bC)$, $G={\rm Spin}(4)$ or $G=\SU(2)$, the latter being the gauge group of the Ashtekar-Barbero formulation of gravity underlying LQG. In this paper we will often assume $G=\SU(2)$.

$\varphi$ is a function on $G^4/G$ where $G$ acts by diagonal right action, \ie\, a function on four copies of $G$ satisfying
\ben
\varphi(g_1,\ldots,g_4)=\varphi(g_1\,h,\ldots,g_4\,h)\quad\forall h\in G.
\label{gauge}
\een
In the geometric interpretation of an elementary GFT quantum as a tetrahedron, $g_I$ define parallel transports along four links through the tetrahedron's faces and (\ref{gauge}) is the invariance under a gauge transformation acting on the vertex where the links meet.

The dynamics is then defined by a choice of action $S[\varphi,\bar\varphi]$ and the functional integral
\ben
Z=\int \mathcal{D}\varphi\;\mathcal{D}\bar\varphi\;e^{-S[\varphi,\bar\varphi]} = \sum_\Gamma \frac{\prod_i \lambda_i^{n_i}}{{\rm sym}(\Gamma)} \,Z[\Gamma]
\een
where $\lambda_i$ are the coupling constants of the theory and the second equality is the perturbative expansion of $Z$ (around $\varphi=0$) in Feynman graphs $\Gamma$, where each $\Gamma$ is associated with a symmetry factor and a Feynman amplitude $Z[\Gamma]$. Each $\Gamma$ forms a two-complex that can be interpreted as a discrete spacetime (with no boundary), or a {\em spin foam}. The amplitude $Z[\Gamma]$ is then the spin foam amplitude defining the dynamics of LQG \cite{LQG3,spinf}; there is a one-to-one correspondence (within a certain class of models) between spin foam models, defined by a choice of $Z[\Gamma]$, and GFT models, defined by a choice of action $S$ \cite{carlomike}.

In addition to the covariant formalism for GFT in terms of the functional integral, there is also a canonical formalism that is closer to canonical LQG \cite{daniele2q}. One defines a Fock space by starting with a Fock vacuum $|\emptyset\rangle$ annihilated by the field operator, $\hat\varphi(g_I)|\emptyset\rangle = 0$, and imposing canonical commutation relations
\ben
[\hat\varphi(g_I),\hat\varphi(g'_I)]=[\hat\varphi^\dagger(g_I),\hat\varphi^\dagger(g'_I)]=0\,,\quad[\hat\varphi(g_I),\hat\varphi^\dagger(g'_I)]={\bf 1}(g_I,g'_I)
\een
as in usual non-relativistic bosonic field theory. ${\bf 1}(g_I,g'_I)$ is an identity element on the space of fields compatible with (\ref{gauge}); for compact $G$
\ben
{\bf 1}(g_I,g'_I):=\int {\rm d}h\;\prod_{I=1}^4 \delta(g'_I h g_I^{-1})
\een
where here and in the following ${\rm d}h$ is the normalised Haar measure on $G$. A one-particle state $|g_I\rangle:=\hat\varphi^\dagger(g_I)|\emptyset\rangle$ is then identified with a geometric tetrahedron, or an open 4-valent spin network vertex in LQG; more complicated simplicial geometries or LQG spin networks correspond to many-particle Fock states. For an explicit construction of this Fock space starting from LQG spin-network wavefunctions, in complete analogy to the usual introduction of second quantisation starting from $N$-particle wavefunctions, see \cite{daniele2q}.

One can use this Fock space construction to define {\em condensates} in analogy with condensed matter physics. Such condensates can be characterised by a non-zero expectation value for the field operator, the GFT field $\hat\varphi(g_I)$, and a large number of particles (potentially infinite) with respect to the Fock vacuum $|\emptyset\rangle$. They are usually defined as superpositions of states with different particle number. In the context of LQG, this means that one is not working on a fixed graph or discretisation of space, but taking a superposition of many discrete geometries into account.

The simplest states used to describe such a condensate are of the form
\ben
|\sigma\rangle:=\mathcal{N}(\sigma)\exp\left(\int ({\rm d}g)^4\;\sigma(g_I)\,\hat\varphi^\dagger(g_I)\right)|\emptyset\rangle\,,
\label{state}
\een
which is a coherent state, an eigenstate of the field operator $\hat\varphi(g_I)$ with eigenvalue $\sigma(g_I)$. As discussed in more detail below, the state (\ref{state}) describes a mean-field approximation where the (non-zero) mean field is given by the `condensate wavefunction' $\sigma(g_I)$.

One now observes that the domain space $G^4/G$ is a finite-dimensional space of geometric configurations, here the possible configurations of a tetrahedron, given in terms of the parallel transports of a (discrete) connection. This space still contains gauge-variant data, as one has to take into account gauge transformations acting from the left (on the open ends of the spin network links). In LQG, the gauge-invariant state is defined on a `dipole', a graph given by four links and two vertices such that all links connect the two vertices, and the space of wavefunctions is $L^2(\SU(2)\backslash \SU(2)^4/\SU(2))$. Accordingly, in (\ref{state}) we require left-invariance of the mean field $\sigma(g_I)$, $\sigma(g_I)=\sigma(hg_I)$ ($\sigma$ is already right-invariant because of (\ref{gauge})). For $G=\SU(2)$, the space $G\backslash G^4/G$ is six-dimensional; a parametrisation of the resulting 12-dimensional phase space in terms of geometric variables corresponding to metric and connection is given in \cite{example}. Viewing the tetrahedron as a locally homogeneous patch of the universe, one can view this phase space as the space of homogeneous (intrinsic and extrinsic) geometric data, \ie\, of homogeneous 3-metrics and Ashtekar-Barbero connections or second fundamental forms.

For the GFT formalism \cite{aristdaniele}, as for LQG \cite{fluxlqg}, there exists a dual `flux representation' which is obtained using a non-commutative Fourier transformation on $G$ defined by \cite{ncfourier}
\ben
\tilde\varphi(B_I) := \int ({\rm d}g)^4\;\left(\prod_{I=1}^4 e_{g_I}(B_I)\right)\,\varphi(g_I)
\een
where $B_I$ are elements of the Lie algebra of $G$, and $e_{g}(B)$ is a choice of plane waves on $G$, satisfying standard properties such as compatibility with the adjoint action of $G$, $e_g(hBh^{-1})=e_{h^{-1}gh}(B)$, and completeness $\int {\rm d}B\;e_g(B)=\delta(g)$. A standard choice for $\SU(2)$ is $e_g(B)=\exp(\frac{\im}{2}{\rm tr}(g B)/\kappa)$ (where $\kappa$ and $B$ have physical dimensions of area). In general, the choice of plane waves corresponds to a (necessary) choice of quantisation map for the fluxes \cite{dmc}, and the flux representation depends on this choice. Functions in the image of the Fourier transform are then equipped with a non-commutative $\star$-product defined by
\ben
e_g(B)\star e_{g'}(B):=e_{gg'}(B)
\een
on plane waves and extended by linearity. In particular, from (\ref{gauge}) one finds that
\ben
\tilde\varphi(B_I) = \int ({\rm d}g)^4\;{\rm d}h\;\left(\prod_{I=1}^4 e_{g_I}(B_I)\star e_{h}(B_I)\right)\,\varphi(g_I) = \tilde\varphi(B_I)\star \delta_\star\left(\sum_I B_I\right)
\een
with the non-commutative delta function $\delta_\star(B):=\int {\rm d}g\;e_g(B)$; the field $\tilde\varphi(B_I)$ satisfies the (non-commutative) {\em closure constraint} $\sum_I B_I=0$. If the $B_I$ are identified with bivectors representing integrals of an area 2-form over the faces, $B^{AB}_I\sim\int_{\triangle_I}e^A\wedge e^B$, the closure constraint means that the faces close to form a tetrahedron. The dual Lie algebra variables (corresponding to `fluxes' of the triad in LQG) hence represent a (discrete) metric, canonically conjugate to the discrete connection given by $g_I$.

The connection to cosmology is made by rewriting the GFT quantum dynamics as dynamical equations for the mean field $\sigma(g_I)$ or its non-commutative Fourier transform $\tilde\sigma(B_I)$ and by constructing macroscopic geometric observables out of the elementary group and Lie algebra variables. Such observables are identified with cosmological variables; for instance, the total volume in a GFT condensate defines the volume of a region of the universe, $a^3 V_0$ if $a$ is the scale factor and $V_0$ the coordinate volume. Such a region can be thought of as the `fiducial cell' in LQC \cite{LQC1,LQC2,LQC3} (which, as in LQC, can make up the whole universe for a compact universe), and itself consists of a large number of fundamental GFT quanta. Constraints between the cosmological variables define the effective cosmological description of the GFT dynamics (an effective `Friedmann equation').

In this paper we make this correspondence between GFT observables and cosmological variables, the crucial step in the physical interpretation of the effective GFT dynamics, more precise. We revisit the previous constructions, mainly given in \cite{condlong,nscaling}, and argue that, in the hydrodynamic approximation, the effective continuum classical geometry reconstructed from a GFT condensate is not necessarily homogeneous. We show how different GFT observables can be used to extract statistical information both about the homogeneous mode and the inhomogeneities. These points are general, and we illustrate them only in the simplest case of an isotropic universe where they are clearest. This simplest case will already be sufficient to make two points: the apparent near-homogeneity of the observed Universe requires a statistical distribution of `patches' that are `large' on average; and for a homogeneous, isotropic universe we give, under a few simplifying assumptions, a derivation of the LQC `improved dynamics' prescription for holonomy corrections. We will see how the presence of inhomogeneities affects the effective cosmological equations.

\section{Effective classical geometry from global GFT observables}
\label{sec:geom}

Part of the geometric information contained in any GFT Fock state, \eg\, a condensate state of the form (\ref{state}), can be expressed in terms of expectation values of suitable second-quantised operators. We focus on `one-body operators' of the general form
\ben
\hat{O} := \int ({\rm d}g)^4\;({\rm d}g')^4\;\hat\varphi^\dagger(g_I)\,O(g_I,g'_I)\,\hat\varphi(g'_I)\,.
\label{operator}
\een
Such operators are extensions of operators $\hat{o}$ of the first-quantised formulation in that the matrix elements of $\hat{o}$ are inserted into the integral (\ref{operator}), $O(g_I,g'_I):=\langle g_I|\hat{o}|g'_I\rangle$ for single-particle states $|g_I\rangle$ and $|g'_I\rangle$. This is the standard procedure in second quantisation; for further discussion of GFT as a second quantised formulation of LQG, see \cite{daniele2q}.

Equivalently, given a representation of the matrix elements of the first-quantised operator $\hat{o}$ in terms of single-particle wavefunctions,
\ben
\langle \Psi|\hat{o}|\Psi'\rangle = \int ({\rm d}g)^4\;\overline{\Psi(g_I)}(\mathfrak{O}\Psi')(g_I)
\een
where $\mathfrak{O}$ is in general a differential operator on $G^4$, (\ref{operator}) can be written as
\ben
\hat{O} := \int ({\rm d}g)^4\;\hat\varphi^\dagger(g_I)(\mathfrak{O}\hat\varphi)(g_I)\,.
\label{operep}
\een
The two expressions are equivalent as $|\Psi\rangle=\int ({\rm d}g)^4\; \Psi(g_I)|g_I\rangle$ in terms of the basis $\{|g_I\rangle\}$.

For example, for $G=\SU(2)$ and interpreting GFT Fock states as LQG spin networks, the one-body operators corresponding to `total fluxes' are defined as
\ben
\hat{b}^i_I := \im\kappa\int ({\rm d}g)^4\;\hat\varphi^\dagger(g_J)\frac{{\rm d}}{{\rm d}t}\hat\varphi\left(\exp\left(\tau^i_I t\right)g_J\right)\Big|_{t=0}
\label{totflux}
\een
where $\kappa$ is an area, the `Planck' area (in kinematical LQG, $\kappa=8\pi\gamma\hbar G_{{\rm N}}$ with the Barbero-Immirzi parameter $\gamma$ and Newton's constant $G_{{\rm N}}$), and $\tau^i$ is a basis of the Lie algebra of $\SU(2)$, usually taken as $\tau^i=\frac{\im}{2}\sigma^i$ in terms of the Pauli matrices $\sigma^i$.

The total area associated to all $I$-th faces in a given state is the expectation value of
\ben
\hat{A}_I:=\kappa\int ({\rm d}g)^4\;\hat\varphi^\dagger(g_J)\sqrt{-\Delta_{g_I}}\,\hat\varphi\left(g_J\right)
\label{areas}
\een
where $\Delta_{g_I}$ denotes the Laplace-Beltrami operator on $\SU(2)$, acting on the $I$-th argument of $\hat\varphi$. The square root is to be defined in terms of eigenvalues of $-\Delta_{g_I}$, in the sense of Dirac \cite{diracbook}.  These eigenvalues are of the form $-j(j+1)$ for integer or half-integer $j$, giving the celebrated discrete area spectrum of LQG. More concretely, introducing the coordinate system on (one half of) $\SU(2)$
\ben
g=\sqrt{1-\vec{\pi}}\,{\bf 1}-\im\vec\sigma\cdot\vec\pi\,,\quad|\vec\pi|\le 1
\label{su2coord}
\een
(which associates to $g\in\SU(2)$ a Lie algebra element $\pi[g]:=\im\vec\sigma\cdot\vec\pi[g]\in\mathfrak{su}(2)$), we have
\bena
\hat{b}^i_I&=&\frac{\im\kappa}{2} \int\left(\frac{{\rm d}\vec\pi}{\sqrt{1-\vec\pi^2}}\right)^4\;\hat\varphi(\vec\pi_I)^\dagger\left(-\sqrt{1-\vec\pi^2_I}\partial_i^I+{\epsilon_i}^{jk}\pi^I_j\partial^I_k\right)\hat\varphi(\vec\pi_I)\,,\nonumber
\\\hat{A}_I&=&\frac{\kappa}{2}\int\left(\frac{{\rm d}\vec\pi}{\sqrt{1-\vec\pi^2}}\right)^4\;\hat\varphi(\vec\pi_I)^\dagger\,\sqrt{-(\delta^{ij}-\pi^i_I\pi^j_I)\partial^I_i\partial^I_j+3\pi^i_I\partial^I_i}\,\hat\varphi(\vec\pi_I)\,,
\label{operators}
\eena
where there is summation over repeated indices $i,j,k$ but not over $I$ which is fixed. If we identify the differential operator appearing in (\ref{operators}) with a first-quantised flux operator $\mathfrak{b}^i_I$ as in (\ref{operep}), \ie
\ben
\hat{b}^i_I =: \int ({\rm d}g)^4\;\hat\varphi^\dagger(g_I)(\mathfrak{b}^i_I\hat\varphi)(g_I)\,,
\een
it is easy to check that indeed
\ben
\hat{A}_I = \int ({\rm d}g)^4\;\hat\varphi^\dagger(g_I)\left(\sqrt{\sum_i\mathfrak{b}^i_I\mathfrak{b}^i_I}\,\hat\varphi\right)(g_I)\,.
\een

Operators of the form (\ref{operator}) or (\ref{operep}) are defined on the whole Fock space, so that quantities such as `the expectation value of the total area' can be associated to any state. For states of cosmological interest, \eg\, a condensate defined by (\ref{state}), such expectation values are identified with cosmological observables such as the scale factor, and relations between expectation values, derived from the fundamental quantum dynamics, become relations between cosmological observables. This allows the derivation of generalised Friedmann equations from the fundamental GFT dynamics, for specific choices of dynamics \cite{nscaling}.

The analysis of \cite{nscaling} focussed on the simplest one-body operators on the GFT Fock space, the total flux (\ref{totflux}) and a `total group coordinate'
\ben
\hat{\Pi}_I := \int ({\rm d}g)^4 \;\vec\pi[g_I]\;\hat\varphi^\dagger(g_J)\hat\varphi(g_J)
\een
which are the analogue of total momentum and `total position' in the GFT context (the latter is in contrast to the physically more meaningful centre-of-mass position \cite{nscaling}). One could, however, also consider other operators, \eg\, the areas (\ref{areas}), or
\ben
\hat{\alpha}_I := \int ({\rm d}g)^4\;\hat\varphi^\dagger(g_I)\left(\sum_i\mathfrak{b}^i_I\mathfrak{b}^i_I\,\hat\varphi\right)(g_I) = \kappa^2\int ({\rm d}g)^4\;\hat\varphi^\dagger(g_J)(-\Delta_{g_I}\,\hat\varphi)\left(g_J\right)\,.
\label{alphaop}
\een
$\alpha_I$ corresponds to the sum of the squared areas associated to each $I$-th face in a general many-particle state. This is a very different quantity from the square of the area (\ref{areas}),
\ben
\hat{A}_I^2 = \int ({\rm d}g)^4\,({\rm d}g')^4\;\hat\varphi^\dagger(g_I)\left(\sqrt{\sum_i\mathfrak{b}^i_I\mathfrak{b}^i_I}\,\hat\varphi\right)(g_I)\,\hat\varphi^\dagger(g'_I)\left(\sqrt{\sum_i\mathfrak{b}^i_I\mathfrak{b}^i_I}\,\hat\varphi\right)(g'_I)\,,
\een
which is not a one-body operator of the form (\ref{operator}). In general, one cannot expect a simple relation between expectation values of $\hat{\alpha}_I$ and $\hat{A}_I$, although in the case of a particularly simple state such as (\ref{state}) there is a relation, given that all expectation values are expressible in terms of integrals involving the mean field $\sigma$.

For the state (\ref{state}), the expectation values of $\hat{A}_I$ and $\hat{\alpha}_I$ are given by
\ben
\langle \hat{A}_I\rangle = \int ({\rm d}g)^4\;\bar\sigma(g_I)\left(\sqrt{\sum_i\mathfrak{b}^i_I\mathfrak{b}^i_I}\,\sigma\right)(g_I)\,,\quad \langle \hat{\alpha}_I\rangle  =  \int ({\rm d}g)^4\;\bar\sigma(g_I)\left(\sum_i\mathfrak{b}^i_I\mathfrak{b}^i_I\,\sigma\right)(g_I)\,.
\een
If $\sigma(g_I)$ was a quantum-mechanical wavefunction, these expectation values would correspond to expectation values $\langle A\rangle$ and $\langle A^2\rangle$ for an operator $A$, and could be used to compute the variance $\langle A^2\rangle -\langle A\rangle^2$ which contains information about the statistics of the observable $A$. But this is {\em not} the right interpretation in second quantisation. $\hat\alpha_I$ is not $\hat{A}_I^2$ and defines an independent observable for general many-particle states. $\sigma(g_I)$ is a mean field on the `minisuperspace' parametrised by $g_I$, describing the collective properties of the coherent state (\ref{state}). $|\sigma(g_I)|^2$ can be interpreted as a {\em classical} number density of classical `patches';
\ben
\langle\hat{\chi}_C\rangle:=\left\langle\int({\rm d}g)^4\;\chi_C(g_I)\hat\varphi^\dagger(g_I)\hat\varphi(g_I)\right\rangle = \int_C ({\rm d}g)^4\;|\sigma(g_I)|^2
\een
gives the expectation value for the number of quanta for which $\{g_I\}$ are in $C\subset G^4$ ($\chi_C$ is the characteristic function of $C$). Note again the difference between first and second quantisation; the analogue of $\hat\chi_C$ in quantum mechanics would correspond to a projective measurement with eigenvalues 0 and 1, whereas the set of eigenvalues of $\hat{\chi}_C$ is $\mathbb{N}_0$.

In this interpretation, the mean field $\sigma(g_I)$ is used to give a classical statistical description of the hydrodynamic approximation, in which a generic condensate state does {\em not} describe a perfectly homogeneous universe, but rather a distribution of patches with different values for geometric quantities such as \eg\, curvature invariants constructed from the $g_I$. An approximately homogeneous universe arises from a sharply peaked mean field.

Let us make this more precise. In \cite{nscaling}, for the simplest case of an isotropic condensate, the identification of GFT expectation values with cosmological observables was
\bena
\langle\hat{f}_I\rangle = \im\kappa\left\langle\int ({\rm d}g)^4\;\hat\varphi^\dagger(\pi[g_J])\frac{\partial}{\partial\pi^I}\hat\varphi(\pi[g_J])\right\rangle & =: & T_I\;a^2\,,\nonumber
\\\left\langle\hat{\Pi}_I\right\rangle = \left\langle\int ({\rm d}g)^4 \;\vec\pi[g_I]\;\hat\varphi^\dagger(g_J)\hat\varphi(g_J)\right\rangle &=:& N\,V_I\,\sin(l_0\,N^{-1/3}\,\omega)\,,
\label{identi}
\eena
where $T_I$ and $V_I$ are dimensionless $\mathfrak{su}(2)\simeq\bR^3$ elements of order one, $a$ is the cosmological scale factor, $\omega$ the spin connection, $N:=\langle \hat{N}\rangle$ the expectation value of the number operator
\ben
\hat{N}:=\int ({\rm d}g)^4\;\hat\varphi^\dagger(g_I)\hat\varphi(g_I)
\label{numbop}
\een
and $l_0$ is a dimensionless number corresponding to a choice of coordinate units. The overall scaling of $\hat{\Pi}_I$ with $N$ implements the observation \cite{nscaling} that holonomies should be given by {\em intensive} observables and the factor $N^{-1/3}$ inside the argument corresponds to a choice of coordinate system in which the condensate as a whole is extended over a fixed coordinate volume, which essentially corresponds to a fixed fiducial volume in LQC. We can introduce a fiducial volume $V_0$ as in LQC and in the following write $l_0=V_0^{1/3}$.

One issue with using (\ref{identi}) as cosmological variables is that one might expect $T_I$ and $V_I$ to be zero by $\SU(2)$ symmetry\footnote{Thanks to Lorenzo Sindoni for pointing this out.} and hence prefer $\SU(2)$ invariant quantities such as the area operator $\hat{A}_I$. But there is another issue with the consistency of the low-curvature limit: as $\hat{f}_I$ and $\hat{\Pi}_I/N$ are canonically conjugate at least approximately as $|\vec\pi|\ll 1$, they should correspond to classical observables with Poisson brackets
\ben
\left\{\sin((V_0/N)^{1/3}\,\omega),a^2\right\}\propto\frac{\kappa}{\hbar}+O(\omega)\,.
\een
But in the limit of small $\omega$ the left-hand side is $(V_0/N)^{1/3}\{\omega,a^2\}$, and hence $\{\omega, a^2\}$ must scale with the number of quanta as $(N/V_0)^{1/3}$, so that the resulting classical Poisson brackets depend on $N$ and hence indirectly on quantities like the scale factor\footnote{We are excluding the possibility that $N$ is a universal constant independent of cosmological observables, which would seem hard to justify physically.}. This seems to be in conflict with the Poisson brackets of classical general relativity for which $\{\omega,a^2\}\propto G_{{\rm N}} V_0^{-1}$ \cite{LQC1}. In order to obtain Poisson brackets consistent with the classical limit at low curvature, the total area must be defined with a nontrivial scaling $N^{1/3}$, and we define
\ben
\langle \hat{A}_I\rangle  = \kappa \left\langle \int ({\rm d}g)^4\;\hat\varphi^\dagger(g_J)\sqrt{-\Delta_{g_I}}\,\hat\varphi\left(g_J\right)\right\rangle =: a^2 N^{1/3} V_0^{2/3}
\label{newident}
\een
with the same $V_0$ as above (the total coordinate volume associated to the condensate), so that $a^2V_0^{2/3}$ is a physical area. Here $I$ is fixed so that we focus on only one of the four areas, as should be sufficient for isotropic universes. For a heuristic picture giving some additional justification to the factor $N^{1/3}$ in (\ref{newident}), one can think of a large `fiducial cell' composed of many elementary cells; one would associate the cosmological area $a^2 V_0^{2/3}$ to the area of one of the sides of this cell. However, the total area $\langle \hat{A}_I\rangle$ does not give just the area of one of these sides but, since one is summing over all tetrahedra, overcounts by a factor $N^{1/3}$. The scaling is also consistent with constructions such as \cite{alesccian} rooted in LQC.

While (\ref{newident}) is simply a definition of $a$, different from the previous one in (\ref{identi}),
also assuming that the expectation value $\langle \hat{A}_I\rangle$ defines an extensive observable (\ie \,$\langle \hat{A}_I\rangle\propto N$) seems to imply that $a^3\propto N$. We will confirm this argument explicitly in section \ref{sec:improdyn}, and derive this relation from an approximation of the GFT dynamics.

We can now introduce additional geometric observables for GFT condensates and, in the hydrodynamic approximation, identify these with cosmological (metric) variables. This identification will make the effective classical statistical distribution of inhomogeneities explicit. For simplicity, we provide this identification for the case of an isotropic condensate (given \eg\, by (\ref{state}) with a choice of $\sigma(g_I)$ that incorporates isotropy). These observables we consider are again expectation values of one-body operators on the GFT Fock space. 

The hydrodynamic approximation to the quantum dynamics replaces the full quantum properties of the state by a finite number of expectation values; it can be viewed as a replacement of the quantum state by a classical statistical distribution of classical particles with properties such as momentum and position in condensed matter physics and areas, angles, etc. in the case of GFT. For instance, in this semiclassical description the condensate has a given number of quanta $N\gg 1$, given by the expectation value of the number operator $\hat{N}$, whereas the quantum state is really a superposition of states with different numbers of particles. We will further clarify this interpretation in section \ref{sec:interp}; for a general discussion of this statistical nature of the hydrodynamic approximation, see also \cite{danielenote}. Here we use the statistical distribution on minisuperspace, given by the mean field, to define further observables that allow us to extract information about cosmological inhomogeneities.

In this effective classical picture obtained from taking expectation values, the mean field defines a statistical distribution of microscopic geometries for $N$ classical patches. For large $N$, we can represent this continuous distribution approximately by a collection of $N$ patches, each labelled by an index ${\bf i}$ and associated with an area $A_I^{{\bf i}}$, such that the statistical distribution of the different values $A_I^{{\bf i}}$ corresponds to the distribution given by the mean field. One can think of the $A_I^{{\bf i}}$ as each representing a sufficiently small region in minisuperspace such that the average number of patches in this region is one. In this approximation, the statistical distribution for a classical random variable $A_I$ is well represented by a single `universe' made up of $N$ patches with definite $A_I^{{\bf i}}$. Within this approximation, (\ref{newident}) becomes
\ben
\langle \hat{A}_I \rangle  = a^2 N^{1/3} V_0^{2/3} = \sum_{\bf i} A_I^{\bf i} =: \sum_{\bf i}  \left(\frac{V_0}{N}\right)^{2/3} a^2\left(1-2\psi^{\bf i}\right)\,,
\label{inhident}
\een
where in the last equality we have defined a quantity $\psi^{\bf i}$ in terms of $A_I^{\bf i}$. To avoid overcounting, the `perturbations' are constrained to satisfy $\sum_{\bf i}\psi^{\bf i}=0$, so that the scale factor $a$ is obtained from the average over the $A_I^{\bf i}$. The factor $(V_0/N)^{2/3}$ in (\ref{inhident}) is the coordinate area associated with each patch. 

(\ref{inhident}) is simply a definition of a geometric quantity $\psi^{\bf i}$. For small $\psi^{\bf i}$ and isotropy, so that metric fluctuations can be captured by a single (volume) variable, one would expect to recover the usual formalism of linear cosmological perturbations \cite{pertrev} from this setting, where one would really think of $\psi$ as a gauge-invariant (Bardeen) potential. At this kinematical level, the $\psi^{\bf i}$ are not necessarily small in any sense. The precise connection between this GFT formalism and the setting of linear perturbations in cosmology, and the physical interpretation of a scale factor $a$ obtained from such an average, will presumably only be clear for condensates for which $\psi\ll 1$ and one can treat $\psi$ as a linear perturbation.

Again, in this approximation in which the statistical distribution over geometric data is approximated by $N$ classical quantities $A_I^{\bf i}$, the expectation value of the operator $\hat\alpha_I$ defined in (\ref{alphaop}) is
\ben
\langle\hat\alpha_I\rangle = \sum_{\bf i} (A_I^{\bf i})^2 = \sum_{\bf i} \left(\frac{V_0}{N}\right)^{4/3} a^4\left(1-2\psi^{\bf i}\right)^2 = \frac{\langle \hat{A}_I \rangle^2}{N} + 4a^4\left(\frac{V_0}{N}\right)^{4/3} \sum_{\bf i} \left(\psi^{\bf i}\right)^2
\een
as $\sum_{\bf i}\psi^{\bf i}=0$. We can now extract information about the inhomogeneities from the expectation values $\langle \hat{A}_I \rangle$ and $\langle\hat\alpha_I\rangle$: for a large number of quanta well approximating a continuum,
\ben
\int {\rm d}^3 x\;\psi(\vec{x})^2\approx\frac{V_0}{N} \sum_{\bf i}\left(\psi^{\bf i}\right)^2=\frac{V_0}{4}\left(\frac{\langle\hat\alpha_I\rangle\, N}{\langle \hat{A}_I \rangle^2}-1\right)
\een
which is expressible only in terms of expectation values of the condensate and $V_0$ which defines a choice of coordinate units.

Similarly, defining an operator $\hat\beta_I^n$ (with $n\ge 3$) by
\ben
\hat\beta_I^n := \int ({\rm d}g)^4\;\hat\varphi^\dagger(g_I)\left(\left(\sum_i\mathfrak{b}^i_I\mathfrak{b}^i_I\right)^{n/2}\hat\varphi\right)(g_I)\,,
\een
we can identify its expectation value with
\ben
\langle\hat\beta_I^n\rangle = \sum_{\bf i} \left(\frac{V_0}{N}\right)^{2n/3} a^{2n}\left(1-2\psi^{\bf i}\right)^{n} = \frac{\langle \hat{A}_I \rangle^n}{N^{n-1}} +\left(\frac{V_0}{N}\right)^{2n/3} a^{2n}\sum_{m=2}^n \binom{n}{m} \sum_{\bf i} \left(-2\psi^{\bf i}\right)^m
\een
and hence (in the continuum approximation) with a weighted sum of all moments of the perturbation $\psi$, of the form $\int {\rm d}^3 x\;\psi(\vec{x})^m$, for $2\le m\le n$. These moments all give independent statistical information on the function $\psi$. In cosmology, $\int {\rm d}^3 x\;\psi(\vec{x})^2$ corresponds to the total power spectrum, whereas the higher moments correspond to the total bispectrum, trispectrum etc., of the scalar perturbations given by $\psi$. These quantities are directly related to cosmological observations (again, in the regime where inhomogeneities are small) which are also statistical in nature. The hydrodynamic approximation in GFT provides us with a classical statistical description of what is really a coherent, homogeneous quantum state of geometry. This idea bears striking resemblance to the mechanism in inflation \cite{kief} of a transition from a quantum state of the inflaton to a classical statistical description for inhomogeneities, which is then observed in the cosmic microwave background. Viewing quantum cosmology as the hydrodynamics of quantum gravity suggests a similar mechanism for quantum geometry \cite{danielenote}.

\section{Homogeneity and classicality conditions for GFT condensates}
\label{sec:interp}

In section \ref{sec:geom} we have argued that an effective classical picture constructed from the hydrodynamic approximation to a generic GFT condensate is to be interpreted as a classical {\em inhomogeneous} universe: the mean field $\sigma(g_I)$ contains statistical information both about the homogeneous mode (corresponding, in the isotropic case, to the scale factor $a$) and about inhomogeneities. In this section, in order to connect with previous work, we recall the discussion of \cite[sec. 3]{condlong} which aimed at constructing states that are candidates for semiclassical, exactly homogeneous universes, and see where it needs to be extended.

The classical phase space variables of a single GFT quantum, interpreted as a geometric tetrahedron, specify a discrete metric and a discrete connection. The momenta conjugate to the group elements $g_I$ are four Lie algebra elements (bivectors) $B_I$, subject to a closure constraint $\sum_I B_I=0$, so that only three are independent. Identifying
\ben
B_i^{AB}={\epsilon_i}^{jk} e_j^A e_k^B\quad(i=1,2,3)
\een
defines a discrete triad $e_j^A$ and a discrete 3-metric by $g_{ij}=e_i^A e_{jA}$ (where indices are contracted with the appropriate $G$-invariant tensor). Equivalently, one can define
\ben
g_{ij}:=\frac{1}{8\,{\rm tr}(B_1B_2B_3)}{\epsilon_i}^{kl}{\epsilon_j}^{mn}\tilde{B}_{km}\tilde{B}_{ln}\,,\quad\tilde{B}_{ij}:=B_i^{AB} B_{jAB}\,.
\label{metrform}
\een
The $g_{ij}$ parametrise the space of gauge-invariant discrete metric data for one tetrahedron. 

In the construction of \cite{condlong}, one considers an embedding of $N\gg 1$ tetrahedra into a manifold in which the coefficients $g_{ij}$ ($i$ and $j$ label the three edges of a tetrahedron emanating from the same vertex) specify a continuum metric expressed in a fixed frame and evaluated at this vertex, \ie
\ben
g_{ij}=:g({\bf e}_i,{\bf e}_j)(x)
\een
where $\{{\bf e}_i\}$ is the given frame and $x$ the position of the given vertex in the embedding. Put differently, in the embedding the edge $i$ is aligned with the vector field ${\bf e}_i$. The construction assumes that the continuum metric to be reconstructed is almost constant on the scale of the tetrahedra, and requires a choice of vector fields ${\bf e}_i$. The latter are fixed by using an embedding into a manifold with topology $\mathcal{M}\simeq \mathfrak{G}/X$ and a transitive group action by the (arbitrary but fixed) Lie group $\mathfrak{G}$. This group action provides a natural choice of $\{{\bf e}_i\}$: take $\{{\bf e}_i\}$ to be a basis of left-invariant vector fields, which is unique up to a choice of scale in the Lie algebra of $\mathfrak{G}$ and a global $\O(3)$ rotation.

It is then clear that a large number of tetrahedra all with the same $g_{ij}$ approximate a spatially homogeneous metric, \ie\, a metric compatible with the group action by $\mathfrak{G}$. 

As a conclusion from this geometric interpretation at the classical level, the following criteria for GFT states to describe spatially homogeneous  classical geometries, {\em at the scale of the GFT quanta}, are given in \cite{condlong}: first, the quantum analogue of the classical property of homogeneity, \ie\, having the same $g_{ij}$ for all tetrahedra, is taken to be `quantum homogeneity', the {\em condensation} of many quanta into the same microscopic quantum state, specified by the mean field $\sigma(g_I)$ in a coherent state such as (\ref{state})\footnote{This {\em quantum} notion of homogeneity, which is crucial to the condensate assumption but must be distinguished from (statistical) homogeneity of the reconstructed classical geometry, is generalised to `wavefunction homogeneity' in the construction of generalised condensate states in \cite{newcon}.}; second, a {\em semiclassicality} condition is required for the `wavefunction' $\sigma(g_I)$, later taken to be the validity of the WKB approximation when applied to $\sigma(g_I)$. Two further conditions --- near-flatness of the geometry on the scale of the tetrahedra, and a large number of quanta $N\gg 1$ --- are required for consistency for the reconstruction procedure and geometric interpretation.

In order to elaborate on the interpretation of these criteria, it may be helpful to recall the physical meaning of the mean-field approximation and the mean field $\sigma(g_I)$ in the context of the physics of Bose-Einstein condensates. This is standard textbook material (and we follow the discussion of \cite[ch.~2]{becbook} closely), which may however be less familiar to practitioners of quantum gravity.

In second quantisation in non-relativistic quantum mechanics, a field operator $\hat\Psi(\vec{r})$ is introduced as a superposition of annihilation operators associated to a complete set of first-quantised wavefunctions,
\ben
\hat\Psi(\vec{r})=\sum_\nu \phi_\nu(\vec{r})\,\hat{a}_{\nu} = \phi_0(\vec{r})\,\hat{a}_0 +\sum_{\nu\neq 0} \phi_\nu(\vec{r})\,\hat{a}_{\nu}\,,
\een
where $\nu$ is a set of labels characterising the states and ``0'' denotes the ground state. In the simplest case of a non-interacting Bose gas in a box, the wavefunction $\phi_0$ is simply a constant and the higher $\phi_\nu$ are plane waves. In the {\em Bogoliubov approximation}, one then replaces the operator $\hat{a}_0$ by the $c$-number $\sqrt{N_0}$, where $N_0$ is the number of atoms condensed into the ground state, and treats the rest as a small fluctuation,
\ben
\phi_0(\vec{r})\,\hat{a}_0 \rightarrow \Psi_0(\vec{r}):=\sqrt{N_0}\,\phi_0(\vec{r})\,,\quad \delta\hat\Psi(\vec{r}):=\sum_{\nu\neq 0} \phi_\nu(\vec{r})\,\hat{a}_{\nu}\,.
\label{bogo}
\een
The mean field approximation is then the limit in which the fluctuations $\delta\hat\Psi$ are ignored. In the words of \cite{becbook}, in this approximation {\em ``[...] the field operator coincides exactly with the classical field $\Psi_0$ and the system behaves like a classical object. This is the analogue of the classical limit of quantum electrodynamics where the classical electromagnetic field entirely replaces the microscopic description of photons.''}

The $N_0$ atoms condensed into the ground state given by $\phi_0(\vec{r})$ are not necessarily found at the same point $\vec{r}_0$, or even close to some $\vec{r}_0$. For the non-interacting gas, they are just evenly distributed over the box. In this case, in momentum space $\phi_0(\vec{p})\propto\delta(\vec{p})$ so that there {\em is} condensation into a single value (zero) for the momentum. This is true for the free gas; for a general interacting system, the ground-state wavefunction will have some finite spread both in momentum and position space, subject to the uncertainty relation
\ben
\Delta x\Delta p \gtrsim \frac{\hbar}{2}\,.
\een
The mean field $\Psi_0$ is not semiclassical in the WKB sense, having constant phase in the simplest case and slowly varying phase more generally.

Coming back to condensates in GFT, we have implemented the mean-field approximation by an appropriate choice of state in the GFT Fock space. The simplest choice (\ref{state}) is simply an eigenstate of the field operator $\hat\varphi(g_I)$ with eigenvalue $\sigma(g_I)$ so that under expectation values (with normal ordering) $\hat\varphi(g_I)\rightarrow \sigma(g_I)$ as in (\ref{bogo}). This is then already the semiclassical description (hydrodynamic approximation) of the full GFT dynamics; the operator $\hat\varphi(g_I)$ is replaced by the classical field $\sigma(g_I)$. 

On the other hand, requiring that the {\em classical} metric geometry one can reconstruct from the hydrodynamic approximation is {\em exactly homogeneous}, at the scale of the GFT quanta, is analogous to requiring the density of the fluid describing a Bose-Einstein condensate in the hydrodynamic approximation to be peaked around a single point (in position space or momentum space). Here we would require that the analogue of the ground-state wavefunction, the mean field $\sigma$ when the GFT dynamics is imposed, is sharply peaked around one particular metric geometry; more concretely, the non-commutative Fourier transform $\tilde\sigma(B_I)$ of $\sigma(g_I)$ has to be sharply peaked around values of $B_I$ that correspond, by (\ref{metrform}), to the same 3-metric $g_{ij}$. This is consistent with our analysis in section \ref{sec:geom}: the magnitude of inhomogeneities depends on the shape of the function $\sigma$ in minisuperspace.  

We can then say that the assumption of condensation is a quantum notion of homogeneity, and the hydrodynamic description arising from the mean-field approximation is a semiclassical approximation\footnote{\label{foot}Generally, outside of the mean-field approximation for coherent states, one can think of other ways of implementing different notions of either homogeneity or semiclassicality, or both. One example is given by the generalised condensates of \cite{newcon}, which are `quantum homogeneous' in the sense of being determined by a single wavefunction for all quanta, but also involve coarse graining of microscopic degrees of freedom. Their hydrodynamic description is not simply a classical limit for the GFT quantum field.}. No further assumption of semiclassicality should be imposed on the mean field, as was already discussed previously  in \cite{example,nscaling}.

Whether there is an approximate classical, homogeneous description of the condensate, in terms of the statistical distribution of patches reconstructed from the hydrodynamic approximation, depends however on the properties of the mean field.

In order to make more precise statements about the magnitude of inhomogeneities, it is necessary to solve the (approximate) GFT dynamics to get a physical $\sigma(g_I)$. There is a fundamental limit on the minimum spread of $\sigma$ in the group and Lie algebra variables. Using the coordinate system on $\SU(2)$ as in (\ref{su2coord}),
\ben
g=\sqrt{1-\vec{\pi}}\,{\bf 1}-\im\vec\sigma\cdot\vec\pi\,,\quad|\vec\pi|\le 1\,,\nonumber
\een
and defining the non-commutative Fourier transform by
\ben
\tilde{f}(B):=\int {\rm d}g\;e^{\frac{\im}{2}{\rm tr}(g B)/\kappa} f(g) = \int\limits_{|\vec\pi|\le 1} \frac{{\rm d}\vec\pi}{\sqrt{1-\vec\pi^2}}\;e^{\im\,\vec\pi\cdot\vec{B}/\kappa}f(\vec\pi)
\label{ftrans}
\een
where $B=:\im\vec\sigma\cdot\vec{B}$ for $B\in\mathfrak{su}(2)$, we have the relation 
\ben
\Delta \pi\Delta B \gtrsim \frac{\kappa}{2}
\label{gftunc}
\een
where we use the requirement on GFT condensates to describe geometries that are nearly flat on the scale of the tetrahedra, \ie\;to be peaked on values $|\vec\pi|\ll 1$ (where $\vec\pi$ are coordinates on a suitable gauge-invariant combination of group elements such as $g_ig_4^{-1}$, see \cite{example}), where (\ref{ftrans}) becomes the standard Fourier transformation. In general, for any $f$, $\Delta\pi\lesssim 1$ which gives a lower bound on $\Delta B$; the image of the non-commutative Fourier transform consists of functions on $\mathfrak{su}(2)\simeq\mathbb{R}^3$ with finite resolution of order $\kappa$.

The uncertainty relation (\ref{gftunc}) has an interesting consequence. The magnitude of the inhomogeneities $\psi$ is controlled by the relative spread $(\Delta B)/\langle B\rangle$. In order to achieve $\psi\sim 10^{-4.5}$ as required observationally, together with near-flatness $\Delta\pi\ll 1$, we then need to have $\langle B\rangle \gg 10^{4.5}\kappa$: the average (physical) length scale associated to the statistical distribution over geometries needs to be at least a few orders of magnitude {\em larger} than the scale set by the `Planck' length $\sqrt{\kappa}$. \footnote{To make this argument quantitative, one needs more insight on the value of $\kappa$ in the microscopic theory. In the standard view of LQG \cite{LQG1,LQG2,LQG3} (see (\ref{totflux}) and below), $\kappa$ is given in terms of the low-energy Newton's constant; if $\kappa$ is affected by renormalisation this would lead to a different fundamental `Planck' length.} Explaining the small magnitude of $\psi$ is one of the basic puzzles of cosmology; here it leads to a condition on physical GFT condensates in order to be observationally viable. The resulting picture is very different from the usual one in LQC where one thinks of elementary geometric quanta of Planck size, corresponding to the lowest non-zero spin $j=1/2$ in LQG \cite{LQCmath}. In the setting of GFT condensates, (\ref{gftunc}) implies that if the average size of the patches is of order $\kappa$, their relative fluctuations resulting in inhomogeneities are at least of order one, so that the universe is very inhomogeneous.

\section{Example: LQC improved dynamics and more}
\label{sec:improdyn}

We have seen in section \ref{sec:geom} how global observables, obtained from expectation values of one-body operators on the GFT Fock space, can be used to statistically distinguish between the homogeneous mode and inhomogeneities. Dynamical equations for such global observables can then be interpreted in cosmological terms. Here we give one example of this in a simple setting that has been studied before \cite{nscaling, condlong, example}, but is now reinterpreted.

The quantum dynamics of a given GFT can be expressed in terms of Schwinger-Dyson equations,
\ben
\left\langle\frac{\delta \mathcal{O}[\varphi,\bar\varphi]}{\delta\bar\varphi(g_I)}-\mathcal{O}[\varphi,\bar\varphi]\frac{\delta S[\varphi,\bar\varphi]}{\delta\bar\varphi(g_I)}\right\rangle = 0
\label{sdeq}
\een
where $\mathcal{O}[\varphi,\bar\varphi]$ is a functional of the GFT field $\varphi$ and its complex conjugate $\bar\varphi$ and $S$ is the action for the GFT model. Choosing $\mathcal{O}=\bar\varphi(g_I)$ and passing to the canonical operator formalism, this becomes
\ben
\left\langle\hat\varphi^\dagger(g_I)\frac{\delta S[\hat\varphi,\hat\varphi^\dagger]}{\delta\hat\varphi^\dagger(g_I)}\right\rangle = 0
\een
with normal ordering under which the delta distribution $\delta\bar\varphi/\delta\bar\varphi$ disappears. Integrating over $G^4$ we obtain
\ben
\left\langle\hat K\right\rangle + \left\langle\int({\rm d}g)^4\,\hat\varphi^\dagger(g_I)\frac{\delta \mathcal{V}[\hat\varphi,\hat\varphi^\dagger]}{\delta\hat\varphi^\dagger(g_I)}\right\rangle = 0
\een
where $\hat{K}$ is the quadratic part of the GFT action and $\mathcal{V}$ contains all higher order terms. The approximation made in \cite{nscaling, condlong, example} is now to neglect the second term. This can be an exact result for certain states and choices of $\mathcal{V}$ \cite{condlong}, or correspond to a weak-coupling limit of the GFT. If we then choose
\ben
\hat{K}=\int ({\rm d}g)^4\;\hat\varphi^\dagger(g_I)\left(\sum_I\Delta_{g_I}+\mu^2\right)\hat\varphi(g_I)
\label{kinetic}
\een
which contains a nontrivial propagator as motivated by studies of GFT renormalisation \cite{renorm1, renorm2}, comparing with (\ref{alphaop}) and (\ref{numbop}) we obtain
\ben
\sum_I\langle \hat\alpha_I\rangle - \kappa^2\mu^2\langle \hat{N}\rangle = 0\,.
\label{dyneq}
\een
Focussing on isotropic universes for simplicity, as in section \ref{sec:geom}, we then identify
\ben
\langle \hat\alpha_I\rangle = \frac{V_0^{4/3}}{N^{1/3}}a^4\left(1+\frac{4}{N}\sum_{{\bf i}}(\psi^{{\bf i}})^2\right)\approx \frac{V_0^{4/3}}{N^{1/3}}a^4\left(1+\frac{4}{V_0}\int {\rm d}^3 x\;\psi(\vec{x})^2\right)
\een
and, assuming all $\langle \hat\alpha_I\rangle$ are equal, the dynamical equation (\ref{dyneq}) reduces to the relation
\ben
a^4\left(1+\frac{4}{V_0}\int {\rm d}^3 x\;\psi(\vec{x})^2\right)\approx \frac{\kappa^2\mu^2}{4}\left(\frac{N}{V_0}\right)^{4/3}
\label{imprel}
\een
between the cosmological variables $a$, $\psi$ and $N$, where the approximation comes from viewing the sum over {\bf i} as approximating a continuum integral over space.

Assuming that the universe is exactly homogeneous, (\ref{imprel}) would imply that
\ben
a^3 V_0 = \left(\frac{\kappa\mu}{2}\right)^{3/2}\, N 
\label{scaling}
\een
which means that for a dynamical state the total physical volume $a^3 V_0$ is proportional to the number of patches $N$, so that each patch has fixed volume $(\kappa\mu/2)^{3/2}$. 

If (\ref{scaling}) holds, the `average group element' $\langle\hat{\Pi}_I\rangle/N$ (see (\ref{identi})) for GFT condensates is identified with the cosmological observable 
\ben
\frac{|\langle\hat{\Pi}_I\rangle|}{N}=\sin\left(\left(\frac{V_0}{N}\right)^{1/3} \omega\right) = \sin\left(\frac{\sqrt{\kappa\mu}}{\sqrt{2}\,a}\,\omega\right)\,.
\label{holo}
\een
The relation (\ref{scaling}) reproduces, in GFT condensate cosmology,  the {\em improved dynamics} prescription of LQG \cite{improv} in which holonomy corrections take the form (\ref{holo}). In order to obtain (\ref{holo}) for GFT condensates, we assumed the form of the kinetic term (\ref{kinetic}), the existence of a weak-coupling limit, and consistency of the Poisson brackets at low curvature that led to (\ref{newident}). Our argument can be contrasted with results in LQC, \eg\,\cite{effdyn} where the improved dynamics scheme was shown to be the only LQC prescription leading, in the (semiclassical) effective dynamics, to a universal `quantum gravity scale' bounding geometric quantities that is independent of initial conditions. The improved dynamics scheme has also been derived from GFT condensate dynamics in \cite{gianl}, using a WKB approximation and the coupling to matter. The two results, obtained using different methods, show different ways of how LQC holonomy corrections can emerge from the more fundamental GFT framework.

For the more general case of inhomogeneous universes, (\ref{imprel}) becomes 
\ben
a^3 V_0\approx \left(\frac{\kappa\mu}{2}\right)^{3/2}\,N\left(1+\frac{4}{V_0}\int {\rm d}^3 x\;\psi(\vec{x})^2\right)^{-3/4}
\label{scaling2}
\een
and hence the total volume $a^3 V_0$ is decreased when inhomogeneities are present.

At the end of section \ref{sec:interp} we have argued that a nearly homogeneous universe is only possible if the length scale associated to the distribution of `patches' is large compared to the `Planck' scale. This statement together with (\ref{scaling2}) would imply that we require $\mu^2\gg 1$ in order to have, within the approximations we are using here, dynamical condensates that are observationally viable, \ie\, satisfy $\psi\ll 1$. This is an example of how input from observation can be translated into constraints on a class of GFT models, as was already advocated in previous works such as \cite{nscaling}. From the perspective of fundamental quantum gravity, (\ref{kinetic}) with $\mu^2\gg 1$ might be viewed as a limiting case in which the Laplace-Beltrami operator is a `small perturbation' to the trivial kinetic term on which existing GFT/spin foam models related to Plebanski gravity are based, \eg\,the class of GFT actions in \cite{carlomike}.

\section{Discussion}
\label{sec:disc}

The `condensate wavefunction' appearing in the mean-field approximation is not a `wavefunction of the universe'. In deriving quantum cosmology models from GFT condensate dynamics, one should use an interpretation that takes the physical interpretation of this mean field into account. As we have argued, one way of clarifying this interpretation is to rewrite the  hydrodynamic approximation of GFT condensates as a classical statistical distribution over the space of geometries. This statistical description then already captures inhomogeneities, and hence this approximate classical picture for GFT condensates generally corresponds to an inhomogeneous universe, even though the underlying quantum state is homogeneous in a precise sense, being given by a coherent state of many quanta. 

One main future task will be to find explicit solutions to the GFT dynamics, in particular ones that describe nearly homogeneous universes, in order to be able to extract clear predictions from GFT condensate cosmology. It may also be necessary to go beyond the mean-field approximation, given that the information about inhomogeneities we can extract is rather limited, corresponding to integrals $\int {\rm d}^3 x\;\psi(\vec{x})^n$ for different $n$.

An important r\^{o}le should be played by solutions that are similar to {\em solitons} that appear in the physics of Bose-Einstein condensates \cite{becbook}. The existence of such solutions depends crucially on the nonlinearity of the Gross-Pitaevskii equation. Similarly, the interaction term in GFT is the main ingredient for creating $4d$ spacetime structure out of $3d$ building blocks \cite{LQG2,GFT1,GFT2}, which also suggests the need for solutions to the full, nonlinear equations. The analogue of the Gross-Pitaevskii equation for GFT condensates, in the simplest approximation where the condensate is defined by (\ref{state}), is just the classical GFT equation of motion. Solutions to these classical equations, for GFT models of interest in quantum gravity, have been the focus of previous work \cite{sol1,sol2,sol3} which mainly aimed at exploring the dynamics of perturbations around a non-trivial background. One might try to choose one of these solutions as a candidate mean field $\sigma(g_I)$ for cosmology. However, the solutions used in \cite{sol1,sol2,sol3} are sharply peaked on `flat' geometries\footnote{See also the closely related work in \cite{gfthydro} using coherent states, where (again in $3d$) imposing the dynamics suggests that states should be sharply peaked on flatness, with large spread in the triad.}. This is natural in the context of \cite{sol1,sol3} of models for $3d$ gravity where solutions to the classical theory, defined by a `BF action' $\int {\rm tr}(E\wedge F[A])$, correspond to flat connections, with conjugate triad 1-form $E$ only determined up to transformations $E\mapsto E+{\rm d}_A\eta$ which leave the action invariant \cite{baezbf}. In $4d$ gravity, however, one requires a distribution with finite spread both in terms of curvature and metric variables (where the spread in the metric, in order to match observation, must be very small), which corresponds to a very different type of classical solution.

In section \ref{sec:improdyn} we gave one example of how to extract information about the GFT dynamics already from the linearised equations. Using the cosmological interpretation of expectation values of GFT Fock space operators given in section \ref{sec:geom}, we have reproduced the {\em improved dynamics} prescription of LQC holonomy corrections from a consistency relation between the total volume and the average particle number which means that $a^3\propto N$. The constant of proportionality depends on the GFT coupling constant $\mu^2$; assuming that the scale associated to the `patches' in the condensate is `large', $a (V_0/N)^{1/3}\gg \sqrt{\kappa}$, as seems required for a nearly homogeneous universe, only models with $\mu\gg 1$ would be viable. 

Beyond the simplest condensates we have considered, one might expect `large' patches to arise as an effective description at a mesoscopic scale, after coarse graining of fundamental GFT quanta. The generalised condensates of \cite{newcon} which involve such a coarse graining (see footnote \ref{foot}) might hence be better suited for describing a realistic cosmology.

The calculations of section \ref{sec:improdyn} are a special case of the general discussion of \cite{nscaling} where, using the variables (\ref{identi}), a generalised Friedmann equation was obtained  that depended on the `atomic number' $N$ which has no classical analogue. Such equations can only be interpreted physically when the `equation of state' $N=N(a)$ is known. Here, under the assumption of exact homogeneity, we find the relation (\ref{scaling}) which also depends both on the cosmological variable $a$ and the particle number $N$ but simply fixes $a$ in terms of $N$ (or conversely), and hence provides the relation $N=N(a)$ but no further information about the dynamics which has to come from elsewhere. In a complete derivation of cosmological dynamics 
from GFT, even in the simplest isotropic, homogeneous case, at least two independent relations are required to provide these two separate ingredients (similar to standard cosmology where the Friedmann equations are supplemented by an equation of state). The first step towards this, a main focus of current work \cite{toappear}, is to discard the approximation in which interactions are neglected and to instead find (approximate) solutions to the full nonlinear GFT equations. This should also shed light on the dynamics of inhomogeneities, which ultimately have to be matched with observational constraints.

\section*{Acknowledgements}

Thanks to Gianluca Calcagni and particularly to Daniele Oriti for many comments and discussions on earlier versions of the paper. The research leading to these results has received funding from the People Programme (Marie Curie Actions) of the European Union's Seventh Framework Programme (FP7/2007-2013) under REA grant agreement n$^{{\rm o}}$ 622339.

\end{document}